\documentclass[pre,floatfix,twocolumn, 10pt]{revtex4} 
\usepackage{graphicx}
\usepackage{amssymb}
\usepackage{natbib}
\usepackage{dcolumn}
\usepackage{bm}
\usepackage{color}
\usepackage[colorlinks=true, linkcolor=blue, citecolor=blue]{hyperref}
\usepackage{subfigure}
\usepackage{graphicx,psfrag,xspace}
\usepackage{color}
\usepackage{amssymb,amsfonts,amsmath}
\usepackage{setspace}

\begin{document}


\author{E. A. Jagla} 
\affiliation{Centro At\'omico Bariloche, Instituto Balseiro, 
Comisi\'on Nacional de Energ\'ia At\'omica, CNEA, CONICET, UNCUYO,\\
Av.~E.~Bustillo 9500 (R8402AGP) San Carlos de Bariloche, R\'io Negro, Argentina}

\title{Discontinuous yielding transition of amorphous materials with low bulk modulus}
\begin{abstract}

The yielding transition of amorphous materials is studied with a two-dimensional Hamiltonian model that 
allows both shear and volume deformations. 
The model is investigated as a function of the relative value of the bulk modulus $B$ with respect to the shear modulus $\mu$. When the ratio $B/\mu$ is small enough, the yielding transition becomes discontinuous, yet reversible. If the system is driven at constant strain rate in the coexistence region, a spatially localized shear band is observed while the rest of the system remains blocked. 
The crucial role  of volume fluctuations in the origin of this behavior is clarified in a mean field version of the model.
\end{abstract}

\maketitle

\section{Introduction}

Yield stress materials have a particular response when submitted to a shear stress. If the applied stress is lower than some critical value, after some initial elastic deformation the material is able to resist any further deformation and it remains rigid. However, if the applied stress is above the critical value the material continues to flow and it can (ideally) remain in ``fluid" state indefinitely. In most cases materials with these properties are structurally amorphous, displaying the rigidity of a solid at low applied stresses, but flowing much like liquids when a sufficiently large stress is applied. Yield stress materials are also known as yield stress fluids\cite{coussot,berthier,nicolas}. The variety of materials with these characteristics is remarkable, ranging (in order from increasing size of elementary constituents) from metallic, polymer and colloidal grasses, through foams and emulsions, to granular solids and granular suspensions.

\begin{figure}
\includegraphics[width=8cm,clip=true]{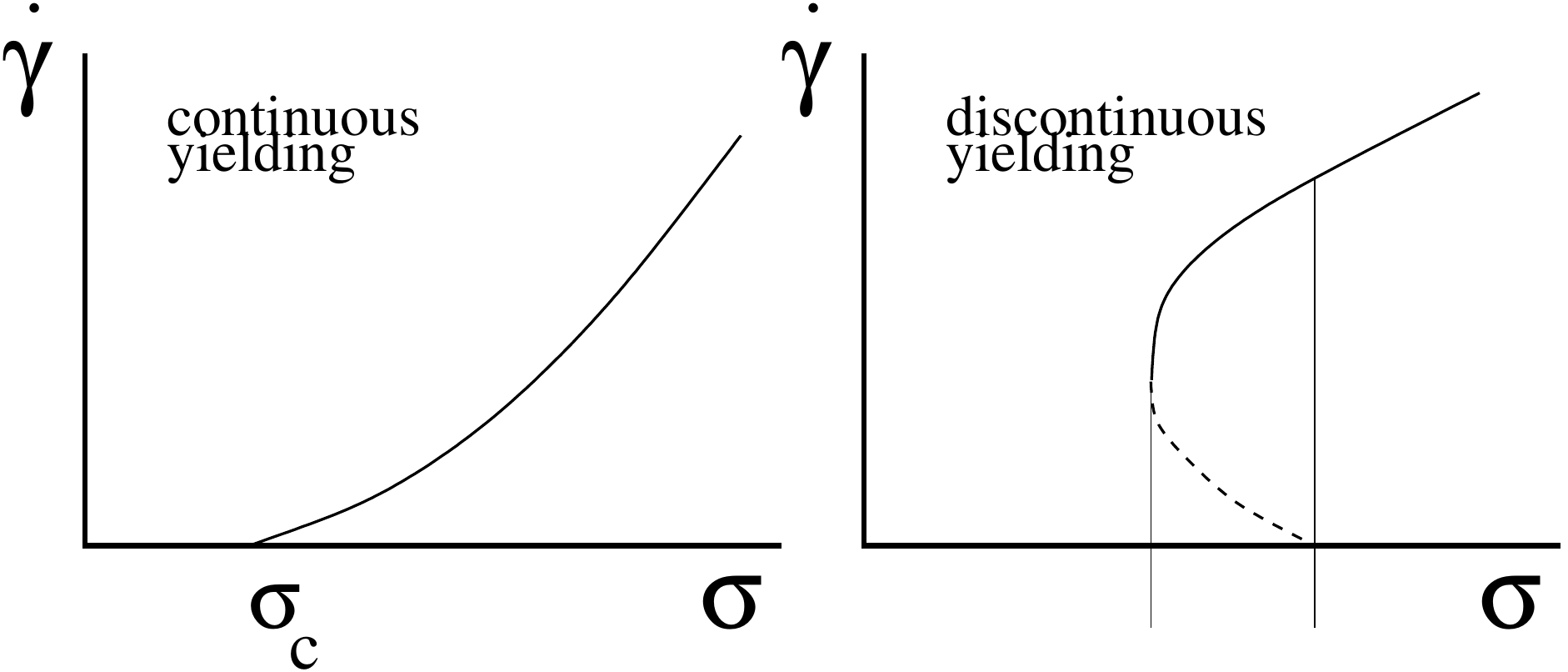}
\caption{Schematic flow curve $\dot\gamma$ vs $\sigma$ for a system exhibiting a continuous (a) or a discontinuous (b) yielding transition}
\label{esquema}
\end{figure}

The stress value that separates the flowing and non-flowing regimes is noted $\sigma_c$, and referred to as the critical stress of the material. Noting by $\dot\gamma$ the values of the deformation rate (or strain rate) in the system, it can thus be stated that $\dot\gamma=0$ if $\sigma<\sigma_c$, and $\dot\gamma>0$ if $\sigma>\sigma_c$. 
In many cases the flow curve (i.e, the value of $\dot\gamma$ as a function of $\sigma$) increases continuously from zero as $\sigma$ is increased passed $\sigma_c$.
When this occurs it is indicated as a ``continuous yielding transition" (Fig. \ref{esquema}(a)) and the material as a ``simple yield stress fluid" \cite{simple_yield_stress_fluids}. The theoretical understanding of this continuous yielding transition has advanced a great deal in the last decades, specially by the recognition of the analogies of this transition (driven by the applied stress) with standard equilibrium thermal phase transitions (driven by temperature)\cite{fisher,kardar}. In fact, one of the most important results obtained was the justification that the value of $\dot\gamma$ increases as $~ (\sigma-\sigma_c)^\beta$ when $\sigma>\sigma_c$. In this expression the value of the ``critical exponent" $\beta$ is named the flow exponent.

Yielding materials that do not behave as ``simple yield stress fluids" are generically classified as ``thixotropic". The precise definition of a thixotropic material has some subtleties, mainly associated to a behavior that is dependent of the shear history of the material. For our purposes however, the important point to realize is that 
thixotropic materials do not reach a state of uniform, stationary stress when uniformly sheared with low values of strain rate \cite{tixo}. Leaving aside its complex non-stationary effects, the main characteristic of a thixotropic material is a reentrance of the flow curve at low strain rate  (Fig. \ref{esquema}(b)), that makes a state of uniform flow at low values of strain rate unstable. In this regime the system separates in a rigid part, and a flowing part in the form of a shear band. This reentrance and coexistence of a flowing and non-flowing regions
can be parallel with the usual coexistence phenomena in first order equilibrium phase transitions.
Sometimes the reentrance and the associated discontinuous transition we are discussing is described as ``discontinuous jamming".

A discontinuous yielding transition  can be originated in a number of reasons.
First of all we stress that we do not refer here to the discontinuous yielding process that occurs for instance in metallic glasses\cite{greer}, where a localized shear band, precursor of sample failure, can be generated upon the application of a sufficiently large stress. Metallic glasses are usually prepared by annealing from the melt, and the state of the material when the shear band appear is well different from the equilibrium state we discuss in the case of thixotropic fluids. Shear bands in metallic glasses are never reversible.

A very robust mechanism that may produce reversible discontinuous yielding and shear banding is referred to as the aging stabilization/strain rejuvenation scenario \cite{picard,olmsted,divoux,jagla_2007}. In this scheme, it is assumed that the sample has some internal mechanism that progressively stabilizes the system increasing its critical stress, as far as the system is at rest. However, if the system is forced to flow, it ``rejuvenates", and its critical stress decreases. This produces an unstable situation in which some part of the system flows and maintains a low critical stress, whereas other parts remain frozen, with a higher critical stress. It was shown that this mechanism in fact produces a flow curve with a reentrance, and a coexistence region between a flowing shear band and a rigid, well aged part of the sample. 
The aging stabilization/strain rejuvenation scenario has also been implemented in simple mean field like models, as for instance in \cite{coussot_2002a,mujumdar,martens_2012}. Although these models justify the reentrant flow curve in the existence of an underlying aging mechanism in the system, the identification of the physical mechanisms of aging remain
mostly an open question\cite{irani}.
 
Another mechanism than can give rise to a discontinuous yielding is referred to as flow-concentration coupling. In suspensions of rigid, dense, non-colloidal particles, \cite{fall_2009,ovarlaz_2006}, and also colloidal glasses \cite{besseling} volume heterogeneities can promote shear banding, since a region of sedimented, percolated particles can coexist with a shear band where particles are shear-induced resuspended.

Independently of its physical origin, in discontinuous yielding a phenomenology similar to that of first
order equilibrium transitions is observed. Particularly, if the system is driven at constant strain rate in the coexistence region, part of the system will be in a flow state, whereas other part will be in a no-flow state. Because of geometric constraints, the flowing part has the structure of a shear band. The strain rate in the band will be the lowest possible value for an homogeneous system, and the band width  adjusts so that the average strain rate in the system is nominally equal to the applied strain rate.

In the present paper we identify another possible mechanism to drive the yielding transition discontinuous. 
We show that when the bulk modulus of the material is sufficiently low, the yielding transition becomes discontinuous and a reversible and stationary shear band appears at low strain rates. \cite{foot}

\section {Numerical modeling}

From the point of view of simulations, continuous yielding has been mainly investigated through atomistic molecular dynamics simulations,\cite{salerno_2021,karmakar} and by mesoscopic models, either ``elastoplastic"\cite{nicolas}, or ``Hamiltonian" \cite{jagla_2007,jagla1,jagla2} models. Molecular dynamics simulations use as input an interaction law between individual particles and other ingredients such as if particles are slippery or sticky, etc. In general, molecular dynamics simulations of point particles find a continuous yielding transition. Elasto-plastic or Hamiltonian modeling use a mesoscopic description of the system in terms of macroscopic coefficients such as bulk and shear modulus that are provided as input parameters. For the sake of simplicity these models are usually written in terms of a single deformation field, namely that corresponding to the applied external deformation. A thorough analysis of the derivation of this kind of models reveals that this typically corresponds to a case in which the bulk modulus is taken to be sufficiently large.
Additionally, simulations have largely concentrated in the   
athermal case, in which thermal fluctuations are neglected, which is typically a realistic assumption due to the relatively large size of the elementary constituents of the physical systems.


Recently, a full mesoscopic scheme based on the consideration of all possible deformations in the system has been presented\cite{tensorial}. This scheme allowed to show for instance that the consideration of a single  deviatoric strain is not sufficient to describe the behavior of a composite sample, in which there are inclusions that are harder, or softer, than the bulk material. Yet in this case the value of the bulk modulus of the material was taken to be very large. 

Here we will investigate within the framework of a Hamiltonian model of yielding, the case in which volume fluctuations are allowed to appear in the system, and in particular in case in which the bulk modulus $B$ of the material is small in such a way that we can expect large volume fluctuations. We will couple these volume fluctuations to a single deviatoric deformation on which the system is externally loaded. Our main result is the finding that the usual continuous yielding transition that occurs if bulk fluctuations are eliminated (when $B\to\infty$) transforms into a discontinuous transition if $B$ is low enough. The flowing pattern in the coexistence region of this discontinuous transition is observed to consist in a band of material that flows, and the rest of the system that remains blocked.

\subsection{Elasto/plastic properties in terms of the strain tensor}

Here we will describe briefly the simulation method for the two dimensional case. A more detailed description is contained in \cite{tensorial}.
Let us consider the (infinitesimal, or linearized) strain tensor $\varepsilon_{ij}$
in terms of the displacement field $u_{i}$

\begin{equation}
\varepsilon_{ij}=\frac 12 \left (\frac{\partial u_i}{\partial x_j}+\frac{\partial u_j}{\partial x_i}\right )
\end{equation}
where $i,j=1,2$. From here we define one volumetric
\begin{equation}
e_1\equiv (\varepsilon_{11}+\varepsilon_{22})/2
\end{equation}
and two deviatoric strains
\begin{eqnarray}
e_2&\equiv& (\varepsilon_{11}-\varepsilon_{22})/2\\
e_3&\equiv& \varepsilon_{12}
\end{eqnarray}
The deviatoric strains are related by a symmetry rotation of $45 \deg$.
Overdamped equations of motion will be used, which are obtained by equating the time derivatives of $e_i$ to (minus) the variation of the total free energy $F$ with respect of $e_i$. In this process it has to be remembered that $e_1$, $e_2$, $e_3$ are not independent, but are related through
\begin{equation}
Q_1e_1+Q_2e_2+Q_3e_3=0
\label{stv2d}
\end{equation}
with 
\begin{eqnarray}
Q_1\equiv\partial^2_x+\partial^2_y\\
Q_2\equiv\partial^2_y-\partial^2_x\\
Q_3\equiv -2\partial_x\partial_y
\end{eqnarray}
This constraint 
follows immediately as an identity after writing $e_1$, $e_2$, $e_3$ in terms of $u_{ij}$. Thus using a Lagrange multiplier $\Lambda$ to satisfy the constraint, the equations of motion are written as
\begin{equation}
\lambda_i \dot e_i=f_i +\Lambda Q_i\\
\end{equation}
where $f_i\equiv -\frac{\delta F}{\delta e_i}$ define the local forces, and $\lambda_i$ are effective viscosity coefficients (note that for further use these coefficients  are allowed to be different for the different deformation modes).
To satisfy the compatibility constraint, we must require (as it is readily verified)
\begin{equation}
\Lambda =-\frac{\sum (f_iQ_i/\lambda_i)}{\sum(Q_i^2/\lambda_i)}
\label{lambda}
\end{equation}

The model is fully defined once we give the form of the free energy $F$.
The free energy will be a spatial integral of a local free energy density, depending on $e_1$, $e_2$, $e_3$.
The dependence on $e_1$ will be simply quadratic, modeling a constant bulk modulus of the material.
The remaining part of $F$ has to encode the fact that the material is amorphous, then $F$ must display a collection of minima,  defining different basins in the $e_2$, $e_3$ plane. 
We define the free energy to be of the form
\begin{equation}
F=\int  \left (\frac B2e_1^2+V(e_2,e_3) \right ) dxdy
\end{equation}
where $V(e_2,e_3)$ are the plastic potential energy, having many minima in the $e_2,e_3$ plane: note that we will consider that the $V$ function is different at different spatial points of the system.

This is the full scheme that was exploited in \cite{tensorial}, in the limit of very large bulk modulus, thus reducing the three coupled equations to two equation for $e_2$ and $e_3$ only. 
Here, we are interested mostly in the coupling between an external deformation with a well defined symmetry (taken to be that of $e_2$) and the volumetric deformation $e_1$. Therefore, the dependence of $F$ on $e_3$ will be assumed to be simply quadratic, and the free energy will be written in the form
\begin{equation}
F=\int  \left (\frac B2e_1^2+V(e_2) +\frac \mu 2 e_3^2 \right ) dxdy
\end{equation}
The equations of motion are explicitly written as
\begin{eqnarray}
\lambda_1 \dot e_1&=&-Be_1+ Q_1\Lambda\label{e01}\\
\lambda_0  \dot e_2&=&f_2(e_2)+ Q_2\Lambda\label{e02}\\
\lambda_0  \dot e_3&=&-\mu e_3+ Q_3\Lambda\label{e03}
\end{eqnarray}
where $f_2\equiv -dV/de_2$.
In these equations we have taken $\lambda_2=\lambda_3\equiv \lambda_0$ (according to the equivalence between $e_2$ and $e_3$ modes). Then $\lambda_1/\lambda_0$ will remain as a free parameter that will allow to consider cases in which the volumetric deformation is more, or less viscous than the deviatoric mode, according to  weather $\lambda_1/\lambda_0$ is greater or smaller than 1.

With the present form of the free energy, the value of $\Lambda$ becomes
\begin{equation}
\Lambda =\frac{\lambda_0BQ_1e_1/\lambda_1-Q_2f_2+\mu Q_3 e_3}{\lambda_0Q_1^2/\lambda_1+Q_2^2+Q_3^2}
\label{lambda2}
\end{equation}
Using the constraint (\ref{stv2d}) to explicitly eliminate $e_3$ from the equations, also taking into account that 
$Q_1^2=Q_2^2+Q_3^2$ it is obtained that the equations of motions for $e_1$ and $e_2$ can be written as
\begin{eqnarray}
\dot e_1&=&-ae_1-b \frac{Q_2}{Q_1}\left (\frac{f_2}{\mu}+e_2\right )\label{e1}\\
\dot e_2&=&f_2-c \frac{Q_2}{Q_1}e_{1}-
d\frac{Q_2^2}{Q_1^2}\left (\frac{f_2}{\mu}+e_2\right )\label{e2}
\end{eqnarray}
where for convenience, the following constants have been defined
\begin{eqnarray}
a&=&\frac{\lambda_0(B+\mu)}{\lambda_1+\lambda_0}\nonumber \\
b&=&\frac{\lambda_0\mu}{\lambda_1+\lambda_0}\nonumber\\
c&=&\frac{\lambda_1\mu-\lambda_0B}{\lambda_1+\lambda_0}\nonumber\\
d&=&\frac{\lambda_1\mu}{\lambda_1+\lambda_0}
\end{eqnarray}

Equations (\ref{e1}) and (\ref{e2}) describe the evolution of the modes with ${\bf q}\ne 0$ of $e_1$ and $e_2$, but 
leave the value of the uniform mode ${\bf q}=0$ undefined. Its evolution is fixed by the driving condition imposed. For a deformation at constant rate $\dot\gamma$ with the symmetry of $e_2$, the uniform mode is set as
\begin{equation}
\overline{e_2}=\dot\gamma t, ~~~\overline{e_1}=0
\label{gdot}
\end{equation}
where the bar indicates average on the whole system, and the stress is calculated as
\begin{equation}
\sigma=\dot\gamma -\overline {f_2}
\label{sigma}
\end{equation}

\begin{figure}
\includegraphics[width=7cm,clip=true]{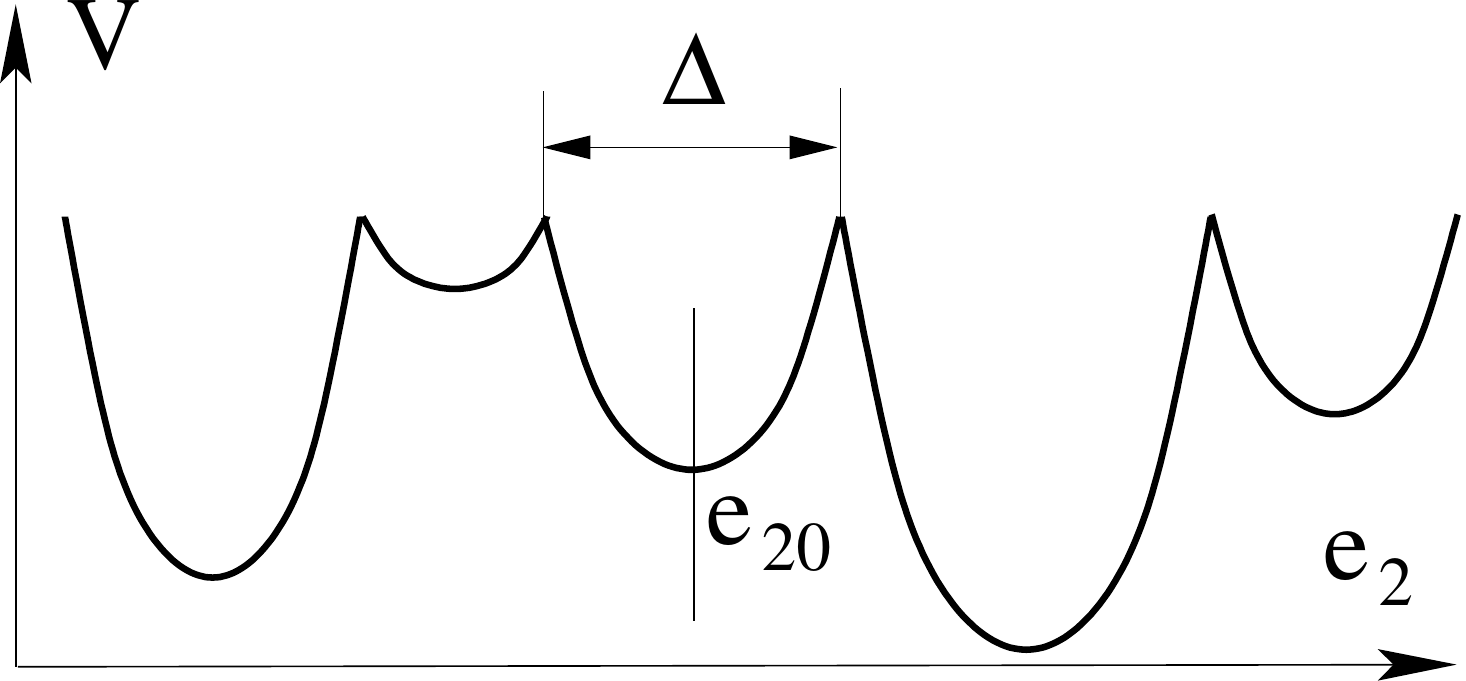}
\caption{A typical realization of the local disorder potential as a function of the deviatoric stress $e_2$. The central value $e_{20}$ and the width $\Delta$ of the parabola are indicated for one particular well.
\label{parabolas}
}
\end{figure}

The final ingredient to completely define the model is to describe the form of the force $f_2$. On one side, as it was stated before, it is considered that this force is different in each spatial position in the system, and totally uncorrelated among different positions. 
In addition, we will consider that $f_2$ is piece-wise linear with respect to a local equilibrium position that we note $e_{20}$, namely $f_2= -\mu(e_2-e_{20})/2$ if $e_2$ is sufficiently close to $e_{20}$. 
However, if $f_2$ becomes too large, a plastic reaccommodation (namely, a variation of $e_{20}$) will occur.
In the end, this situation corresponds to have a contribution from $e_2$ to the free energy and to the force $f_2$ as qualitatively depicted in Fig. \ref{parabolas}. The width $\Delta$ of each parabolic basin is taken to be a stochastic variable with a flat distribution between 0.3 and 1.3. The evaluation of the potential is done on the fly: given some value for $e_{20}$, $f_2$ is calculated as $f_2=-\mu(e_2-e_{20})$ as long as $|e_2-e_{20}|<\Delta/2$. When this last relation is violated,
a new value $\Delta^{new}$ of $\Delta$ is chosen, and also a new value ${e_{20}^{new}}$ of ${e_{20}}$ according to
\begin{equation}
{e_{20}^{new}}={e_{20}}+\Delta/2+\Delta^{new}/2
\end{equation}

\section{Numerical Results}

We simulate Eqs. (\ref{e1},\ref{e2}) using a standard first order Euler method (with a time step of 0.1). We fixed the value of $\dot\gamma$ by making $\overline {e_2}$ satisfy at every step the relation (\ref{gdot}), and $\sigma$ is obtained from Eq. (\ref{sigma}). We start the simulation at the largest values of $\dot\gamma$, and reducing it progressively along the simulation. This guarantees that the sample is well equilibrated and no further concern appears about the initial state of the sample.

\begin{figure}
\includegraphics[width=8cm,clip=true]{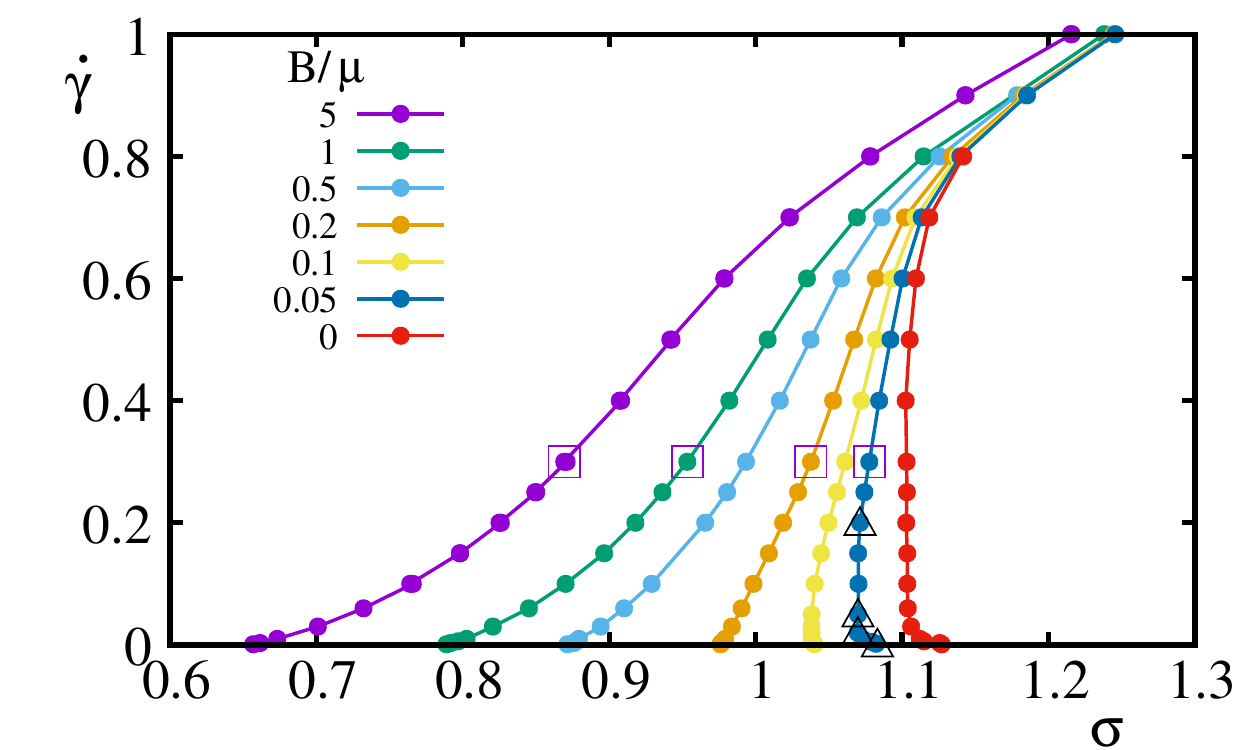}
\caption{Flow curves $\dot\gamma$ vs $\sigma$  at different values of $B/\mu$, as indicated. Simulations were run by fixing $\dot\gamma$ and measuring the average $\sigma$. For $B/\mu\lesssim 0.1$ and low $\dot\gamma$ there is a spatial separation in the system between a non-flowing part, and a flowing shear band. Triangles (squares) highlight  the points corresponding to the panels in Fig. \ref{coexistencia} (\ref{fig_e1}).
\label{resultados1}
}
\end{figure}

\begin{figure}
\includegraphics[width=8cm,clip=true]{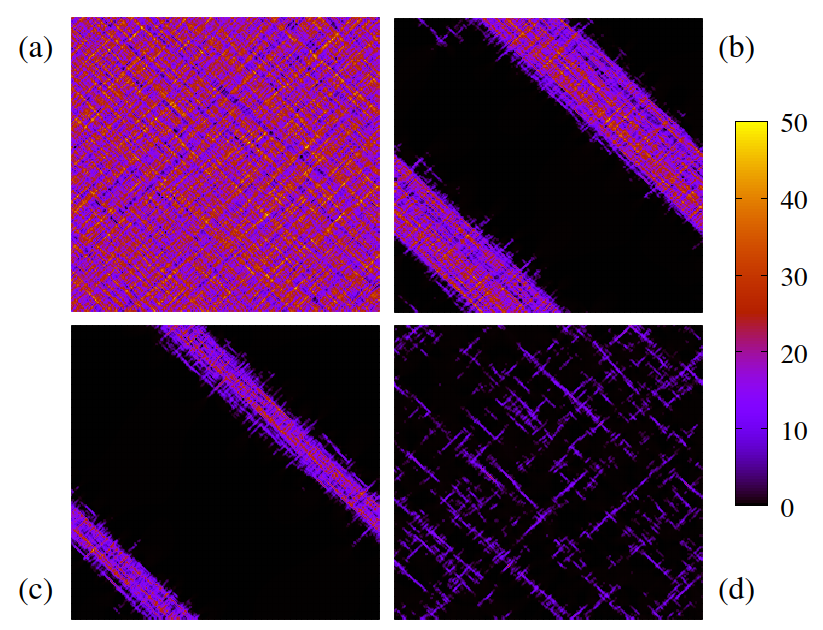}
\caption{Accumulated strain for $B/\mu=0.05$ and values of $\dot\gamma=0.2$, 0.05, 0.02 and 0.001. ($\Delta t=100$ for a,b,c, $\Delta t=1000$ for d).  
\label{coexistencia}
}
\end{figure}

In Fig. \ref{resultados1} we see the form of the flow curve $\dot\gamma$ vs. $\sigma$ obtained in simulations of a system of 256$\times$256 sites, with $\lambda_1=\lambda_0$, and for different values of $B/\mu$. 
We see that reducing the value of $B$ produces an increase of the values of $\sigma$ for a given value of $\dot\gamma$. When $B/\mu$ is sufficiently low ($B/\mu\lesssim 0.15$) the plot acquires a vertical slope, and even a reentrance 
at the lowest values of $\dot\gamma$ is observed. This is evidence of an instability in the homogeneous deformation of the system, that in this region becomes spatially separated between flowing and non-flowing regions. In fact, 
in Fig. \ref{coexistencia} we plot the local accumulated strain over some period of time, at different values of $\dot\gamma$ of the curve corresponding to $B=0.05 \mu$. For $\dot\gamma=0.2$ we can see that the whole sample participates of the deformation, and the local deformation rate in the long run is uniform in all the system. For values of $\dot\gamma=0.05$, and $0.02$ however, we clearly see a region in the form of a band where all deformation accumulates, while region outside the band has essentially zero deformation. This reveals the existence of a region of coexistence in the flow of the system. Note that
the strain rate within the band is uniform, and the band width accommodates to produce a  value of the average strain rate in the system that coincides with the externally imposed value. This is nothing but the phenomenology that is observed in classical first order phase transitions, as for instance in the liquid-gas coexistence. 

\begin{figure}
\includegraphics[width=8cm,clip=true]{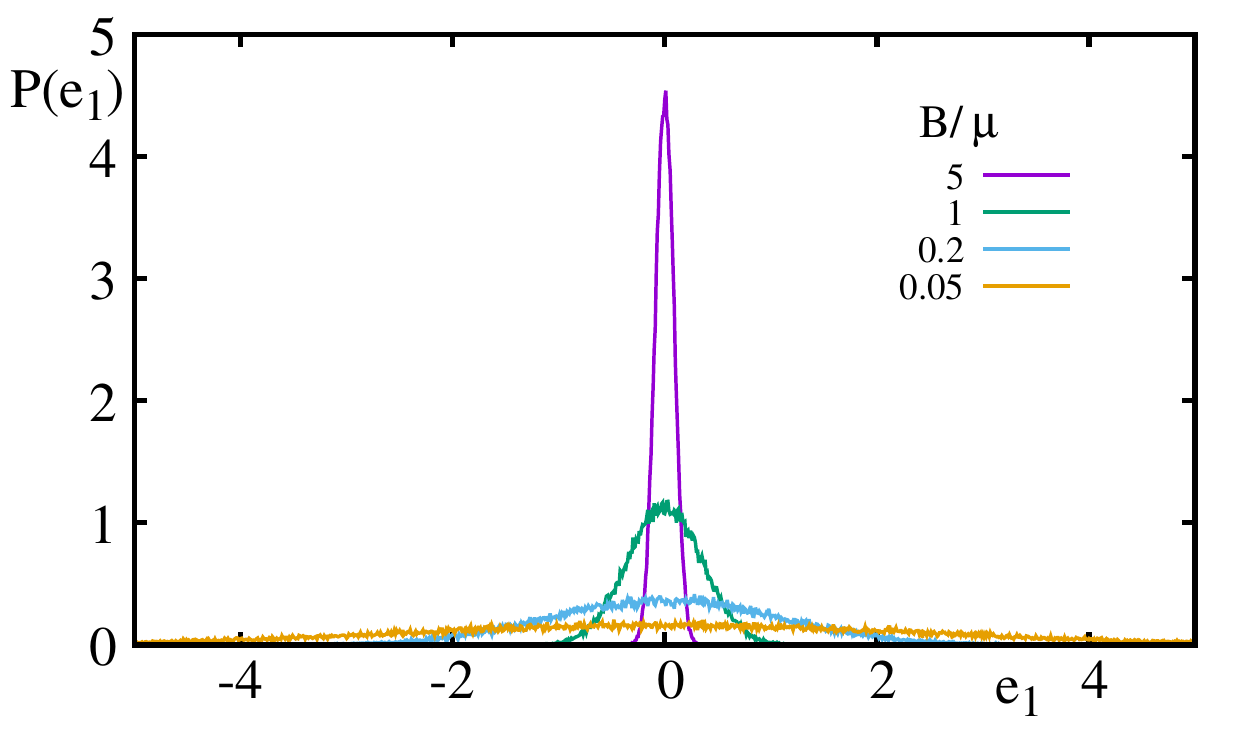}
\caption{Histogram of the distribution of $e_1$ values, at the point marked with squares in Fig. \ref{resultados1} ($B/\mu$ as indicated, $\dot \gamma=0.3$. Note that for this value of $\dot\gamma$ the sample is shearing uniformly). The fluctuations of $e_1$ clearly increase as $B$ is reduced.
\label{fig_e1}
}
\end{figure}

\begin{figure}
\includegraphics[width=6cm,clip=true]{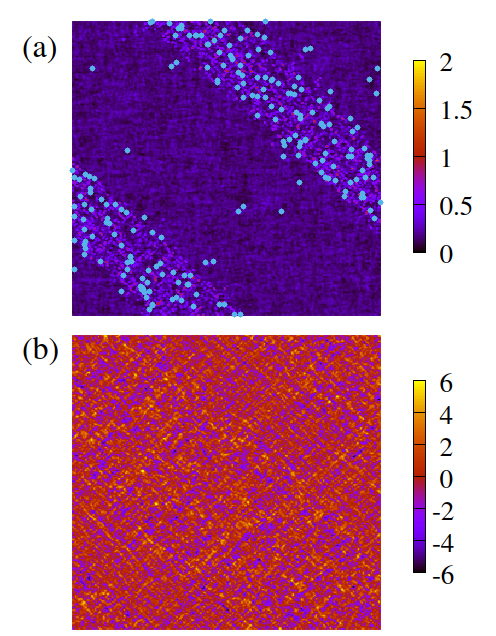}
\caption{(a) Spatial distribution of $\Omega\equiv \Delta-(e_2-e_{20})$, representing the stability range of each site before suffer a plastic deformations, at $B/\mu=0.05$, $\dot\gamma=0.05$. The smallest values of $\Omega$ ($\Omega<0.01$) are highlighted with light blue dots, and  appear mainly within the shear band. 
(b) Spatial distribution of $e_1$ at the same configuration of panel (a). No clear information about the localization of the shear band is obtained from the instantaneous values of $e_1$.
\label{omega}
}
\end{figure}

The formation of the shear band is clearly associated to the low value of the ratio $B/\mu$. When $B$ is reduced the fluctuations in the value of $e_1$ increase. This can be seen in the snapshots shown in Fig. \ref{fig_e1}, where the spatial distribution of $e_1$ is shown at a fixed value of $\dot\gamma$ for samples with different values of $B/\mu$. 
The spatial fluctuation of $e_1$ increases as $B$ is reduced. Yet, notice that the spatial average remains in all cases at the value $\overline{e_1}=0$, that corresponds to the minimum value of the energy. Beyond the clear enhancement of fluctuations of $e_1$ observed at lower values of $B$, it is difficult to find a clear signature of how a reduction of $B$ induces an inhomogeneous flow at low values of $\dot\gamma$. Actually, even for values of the parameters for which a shear band is clearly established in the system, a snapshot of the values of $e_1$ does not reveal any clear difference between points within the shear band and those in the frozen region outside it (Fig. \ref{omega}(a)). A clearer understanding of the origin of the instability leading to the formation of a shear band will be obtained with the analysis of the next section. But we may notice here that a clear difference between the structure of the material within the shear band and outside it is obtained by plotting the values of $\Omega\equiv \Delta-(e_2-e_{20})$ across the system (Fig. \ref{omega}(b)). $\Omega$ represents the additional strain that each site must receive in order to become unstable and experience a plastic deformation. 
The minimum values of $\Omega$ appear within the shear band, namely in the region already flowing, whereas outside it there is a sort of gap $\Omega_0$, in such a way that $\Omega>\Omega_0$. This guarantees that the system will continue to localize the plastic deformation within the shear band.

Coming back to the results in Figs. \ref{resultados1} and \ref{coexistencia}, an interesting phenomenon
is observed at very low strain rates.
We have seen that the local strain rate in the band remains constant and finite, and the width of the band decreases 
as the applied strain rate is reduced. However,
the shear band cannot be arbitrarily thin, since the discreteness of the system imposes a minimum value for its width. In this limit the phenomenology of the flow is the following. The band width stays at its minimum, compatible with the discreteness of the system. However, the band is not persistent in time, it has only a limited time duration, and when reappears it can do so essentially at any position in the sample. This is what can be seen in panel (d) of Fig. 
\ref{coexistencia}. Thus in this regime of very small strain rate the uniform deformation of the sample is effectively 
recovered at very large time intervals, when the fluctuating and extremely thin band has wandered around the whole system.  

\section{Mean field description}

In this section we derive from the model equations (\ref{e1},\ref{e2}) a description in terms of the field $e_2$ alone, and then a mean field description that will
help us to understand more clearly the origin of the discontinuous transition that we observe
in the case in which the bulk modulus is small enough.

The two first order equations (\ref{e1},\ref{e2}) describe a coupled evolution of the volumetric ($e_1$) and deviatoric 
($e_2$) degrees of freedom. 
Before solving for $e_1$ and obtaining a model in terms of $e_2$ alone in the general case, we notice that there are some case where an independent evolution of $e_2$ emerges automatically. The most obvious case corresponds to $B=\lambda_1\mu/\lambda_0$ (i.e., $c=0$), where we see that the term in $e_1$ vanishes in Eq. (\ref{e2}), that then becomes  
\begin{equation}
\dot e_2=f_2-
d\frac{Q_2^2}{Q_1^2}\left (\frac{f_2}{\mu}+e_2\right ) ~~~~~~(B=\lambda_1\mu/\lambda_0)
\label{bmu1}
\end{equation}
Another case corresponds to $B\to\infty$. In this case the time dynamics of $e_1$ is much more rapid than for $e_2$. Thus $\dot e_1$ can be taken as zero in Eq. (\ref{e1}), and the obtained form of $e_1$ replaced in Eq. (\ref{e2}), providing
\begin{equation}
\dot e_2=f_2-
\mu\frac{Q_2^2}{Q_1^2}\left (\frac{f_2}{\mu}+e_2\right ) ~~~~~~(B/\mu\to\infty)
\label{binfty}
\end{equation}
which is identical to (\ref{bmu1}) with a different numerical pre-factor in the last term.
Both Eqs. (\ref{bmu1}) and (\ref{binfty}) admit a very simple mean field analysis. In fact the operator ${Q_2^2}/{Q_1^2}$ has the form (in Fourier space)
\begin{equation}
\frac{Q_2^2}{Q_1^2}=\frac{(q_x^2-q_y^2)^2}{(q_x^2+q_y^2)^2}
\label{operador}
\end{equation}
and therefore can be written in terms of the angle $\theta$ of the $q_x$,$q_y$ plane as $2\cos(2\theta)^2$. The mean field description corresponds to take the angular average of this operator, that is simply equal to $1/2$. In this way, the mean field version of Eq. (\ref{binfty}) reads
\begin{equation}
\dot e_{2{\bf q}}=
f_{2{\bf q}}-  \frac{\mu}{2}\left (\frac{f_{2{\bf q}}}\mu +e_{2{\bf q}}\right )
\label{q}
\end{equation}
The sub-index ${\bf q}$ was added here as a reminder that this equation describes the evolution of the different modes with ${\bf q}\ne 0$. However, since there are no ${\bf q}$-dependent operators in Eq. (\ref{q}), when supplemented with the conditions for the uniform mode (Eqs. (\ref{gdot}), (\ref{sigma})) it is also the equation that describes the evolution of a generic site in the system. Therefore, adding the uniform mode we obtain for the $e_2({\bf r})$ field the equation
\begin{equation}
\dot e_2= f_2-  \frac{\mu}{2}  \left (\frac{f_2}\mu+e_2-
\frac{\overline{f_2}}\mu-\overline{e_2}\right )+\dot\gamma-\overline{f_2}
\label{e2mf}
\end{equation}
The meaning of the average values $\overline {f_2}$, $\overline {e_2}$ is that of the average over the extended system. Using Eq. \ref{gdot}, $\overline {e_2}$ can be replaced by $\dot\gamma t$. 
In order to obtain a mean field equation, applying to a generic site in the systems, we can replace the spatial average $\overline {f_2}$ (note that this is a time-invariant quantity) by the temporal average for the generic site. We note this temporal average as $\langle f_2\rangle$, 
and obtain
\begin{equation}
\dot e_2= f_2-  \frac{\mu}{2}  \left (\frac{f_2}\mu+e_2-\dot\gamma t\right )-
\frac{\langle f_2\rangle}2 +\dot\gamma
\label{mean_field}
\end{equation}
This is a stand alone equation representing a mean field approximation to the original extended problem. It can be numerically integrated to obtain the value of $\sigma$ in the system using Eq. (\ref{sigma}).
The easiest way to do so is to define a shifted variable $\tilde e_2\equiv e_2+
(\langle f_2\rangle-2\dot\gamma)/(2\mu)$, in such a way that Eq. (\ref{mean_field}) becomes
\begin{equation}
\dot {\tilde e}_2= \frac{f_2}{2}+  \frac{\mu}{2}(\dot\gamma  t-\tilde e_2)
\label{ttilde}
\end{equation}
The form (\ref{ttilde}) is particularly convenient as it is exactly the form of the equation describing the Prandtl-Tomlinson (PT) model of friction, that was already introduced in the context of yielding in \cite{jagla2,pt1,pt2}. Eq. (\ref{ttilde}) can be easily simulated, and the value of $\sigma$ at each applied value of $\dot\gamma$ is obtained from Eq. (\ref{sigma}). This value of $\sigma$ can be slightly re-expressed using 
equation (\ref{ttilde}), and be written as \cite{nota2}
\begin{equation}
\sigma= \mu(\dot\gamma t-\tilde e_2)-\dot\gamma
\label{sigma2}
\end{equation}

The flow curve that is obtained from Eqs. (\ref{ttilde})-(\ref{sigma2}), although slightly different from the one in a traditional PT model \cite{nota2} does not display the reentrance that is necessary to justify a spatial separation between a flowing and a non-flowing region in the full model. In order to obtain a mean field model with a reentrance in the flow curve, we must analyze more closely the starting equations (\ref{e1}), (\ref{e2}).
Eq. (\ref{e1}) is a first order, linear differential equation for each ${\bf q}$-mode of the variable $e_1$, and therefore it can be explicitly integrated to obtain
\begin{equation}
e_{1{\bf q}}(t)=-b\int_{-\infty}^t e^{-a(t-\tau)} \frac{Q_2}{Q_1}\left (
\frac{{f_{2{\bf q}}(\tau)}}{\mu}+{e_{2{\bf q}}(\tau)}\right )d\tau
\label{solucion_e1}
\end{equation}
Introducing this expression into Eq. (\ref{e2})
we obtain the following integro-differential equation for $e_{2{\bf q}}$
\begin{eqnarray}
\dot e_{2{\bf q}}=f_{2{\bf q}}-\frac{Q_2^2}{Q_1^2}\left[
\left( \frac{f_{2{\bf q}}}{\mu}+e_{2{\bf q}}\right )d-\right .\nonumber\\
\left .
-bc \int_{-\infty}^t e^{-a(t-\tau)}
\left( \frac{f_{2{\bf q}}(\tau)}{\mu}+e_{2{\bf q}}(\tau)\right )
d\tau
\right]
\end{eqnarray}
Comparing with expressions (\ref{bmu1}) or (\ref{binfty}) we see that the effect of $e_1$ (represented by the integral term) is to reduce the strength of the coupling between different sites in the system. As we will see this reduction is stronger at low values of $\dot\gamma$. As a reduced coupling increases the stress in the system, a reentrant flow curve can emerge from this mechanism. Let us consider as before the mean field case in which $Q_2^2/Q_1^2$ is replaced by its angular value of 1/2. Following the same steps that led to Eq. (\ref{mean_field}) we obtain for the generic mean field site the equation
\begin{eqnarray}
\dot e_2=f_2-\frac d2\left(\frac{f_2}{\mu}+e_2-\dot\gamma t\right )
+\nonumber\\
+\frac{bc}2 \int_{-\infty}^t e^{-a(t-\tau)}
\left(\frac{f_2(\tau)}{\mu}+e_2(\tau)-\dot\gamma \tau\right )
d\tau
+\nonumber\\
+\left(\frac{da-bc}{2a\mu}-1 \right) \langle f_2\rangle +\dot\gamma
\label{full_mean_field}
\end{eqnarray}
where the constant terms were added, as before,
in order to satisfy Eq. (\ref{sigma}).

Before presenting numerical results on Eq. (\ref{full_mean_field}), we highlight two limiting cases.
The typical time scale on which $f_2(\tau)$ and $e_2(\tau)-\dot\gamma \tau$ vary appreciably corresponds to the time in which the external driving sweeps over one period of the energy landscape, namely $\sim \dot\gamma^{-1}$. The integral in Eq. (\ref{full_mean_field})
averages these quantities over a time $\sim a^{-1}$. If $\dot\gamma\gg a$ the integral is $\simeq \langle f_2\rangle/a$, and the equation becomes 
\begin{equation}
\dot e_2=f_2-\frac d2\left(\frac{f_2}{\mu}+e_2-\dot\gamma t\right )
+\left(\frac{d}{2\mu}-1 \right) \langle f_2\rangle +\dot\gamma
\label{large}
\end{equation}

On the other hand, if $\dot\gamma\ll a$, the integral in Eq. (\ref{mean_field}) gives $\simeq (f_2(t)/\mu+e_2(t)-\dot\gamma t)/a$, and the evolution equation becomes
\begin{equation}
\dot e_2=f_2-\frac {da-bc}{2a}\left(\frac{f_2}{\mu}+e_2-\dot\gamma t\right )
+\left(\frac{da-bc}{2a\mu}-1 \right) \langle f_2\rangle +\dot\gamma
\label{small}
\end{equation}
These two equations, corresponding to the limits of very large and very small $\dot\gamma$, have a form equivalent to Eq. (\ref{mean_field}), and therefor display limiting behavior that can be described as realizations of the PT model.

\begin{figure}
\includegraphics[width=7cm,clip=true]{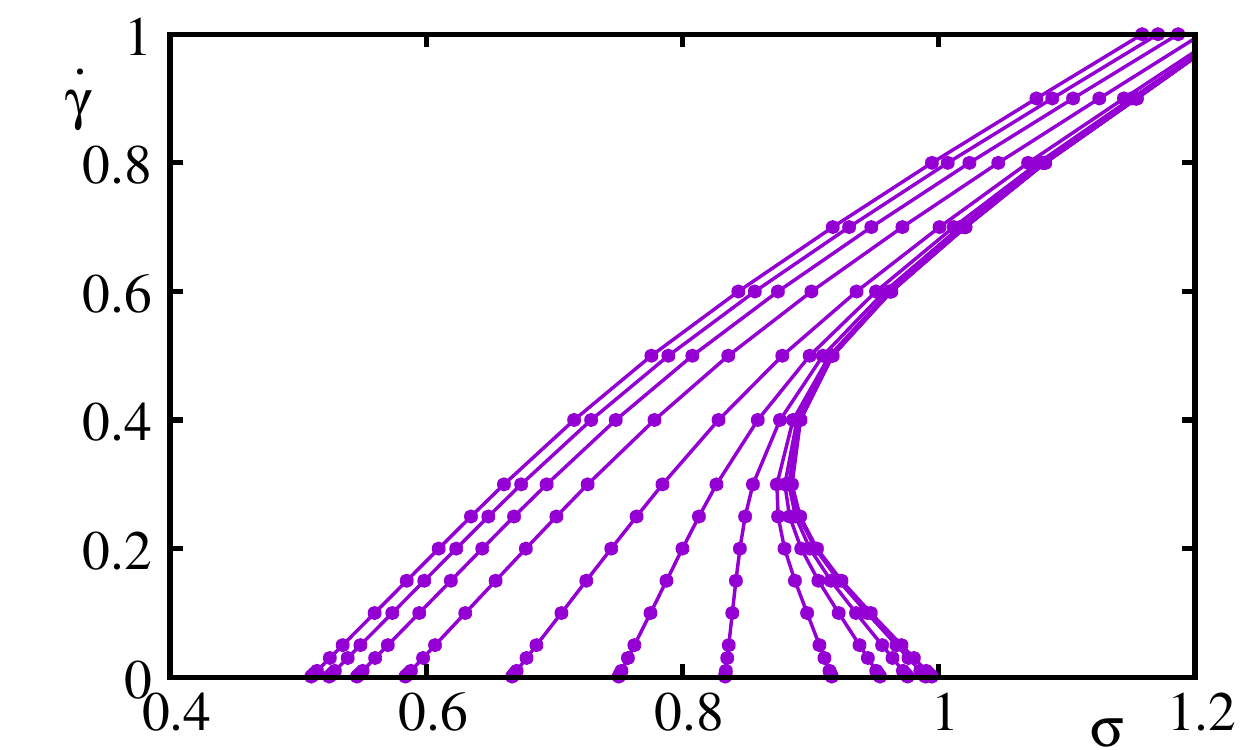}
\includegraphics[width=7cm,clip=true]{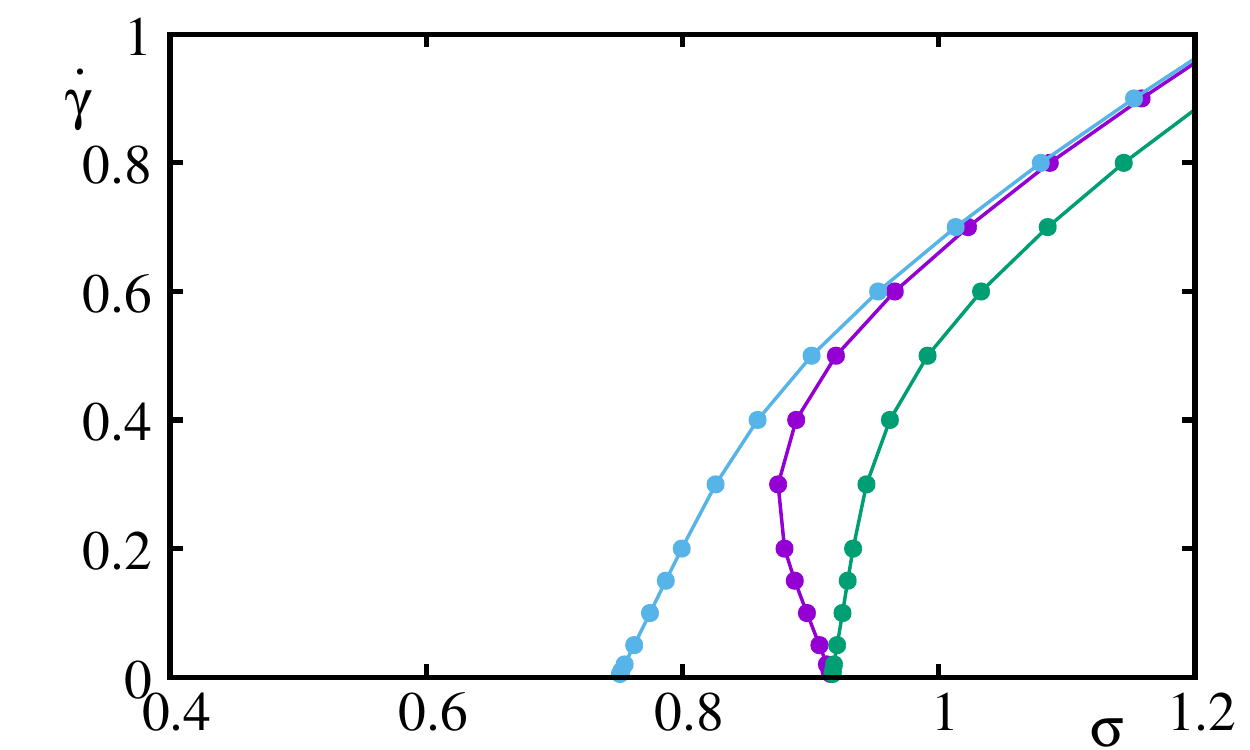}
\caption{(a) Flow curves of the mean field model (Eq. (\ref{full_mean_field})), for different values of $B/\mu$ ($B/\mu$= 50, 20, 10, 5, 2, 1, .5, .2, .1, .05 from left to right). (b) The curve corresponding to $B/\mu=0.2$, and the results of simulating Eqs. (\ref{large}) (light blue) and (\ref{small}) (green) for the same value of $B/\mu$. They display the correct limit at large, and small $\dot\gamma$ respectively, but only the result from Eq. (\ref{full_mean_field}) displays the reentrance at finite values of $\dot\gamma$.
}
\label{flow_curves_mf}
\end{figure}

In Figure \ref{flow_curves_mf}(a) we see the result of numerically simulating Eq. (\ref{full_mean_field}), for the case $\lambda_1=\lambda_0$, and different values of the ratio $B/\mu$.
We clearly see how the value of $\sigma$ increases for the same $\dot\gamma$ as $B/\mu$ is reduced, generating a reentrance when $B/\mu\lesssim 0.3$.
For the case  $B/\mu=0.2$ we also show in Fig. \ref{flow_curves_mf}(b) the result of simulating Eqs. (\ref{large}) and (\ref{small}) that are expected to give an accurate result for large and small values of $\dot\gamma$, respectively. 
We see that in fact this is the case, but notice that none of these two curves is reentrant, meaning that the reentrance effect depends crucially on the progressive softening that the coupling to $e_1$ produces as $\dot\gamma$ is reduced.
It must be pointed out that since this is a mean field model there is no place for the true spatial separation of a flowing and a non-flowing region in the system, and a true reentrance is observed, similar 
to the behavior of the van der Waals model for liquid-gas transition, instead of the coexistence line of the full spatial model.

\label{section:2d}

\section{Conclusions}

We have studied a mesoscopic model for the yielding transition of an amorphous system.
The system is driven by the application of a global external strain rate, and the stress in each part of the system is calculated along the simulations. The flow curve (strain rate vs stress global average value) is one of the main output of the simulation.
The main new ingredient of the present implementation is to incorporate the possibility of 
volume fluctuations through a finite value of the bulk modulus in the system. The amplitude of volume fluctuations is mainly controlled by the relation $B/\mu$
between the bulk and shear modulus of the material. When this ratio is large, volume fluctuations are suppressed, and the model behaves as reported in previous publications, in particular displaying a continuous transition between a non-flowing regime ($\dot\gamma=0$ for $\sigma<\sigma_c$) and a flowing regime  ($\dot\gamma>0$ for $\sigma>\sigma_c$). The possibility of volume fluctuations as $B/\mu$ is reduced produces an increase of the observed value of $\sigma$ at the same $\dot\gamma$. Most importantly, this increase is larger for lower values of $\dot\gamma$ producing (if $B/\mu$ is small enough) a reentrance in the flow curve of the system. This reentrance is the signature of a discontinuous yielding: if the system is driven by controlling $\sigma$, there is a discontinuous jump between $\dot\gamma=0$ and some finite $\dot\gamma\ne 0$ at a particular value of $\sigma$. If instead the system is driven by fixing the average value of $\dot\gamma$, then the spatial distribution of the deformation has a coexistence region, where some part of the system is stuck, and other part is yielding. The geometric constraints implied by the elastic couplings in the system produce that this yielding region is actually a well defined band, namely a shear band in the material.
We have been able to derive a mean field description of the problem in terms of a single shear degree of freedom and showed in this case how the effect of volume fluctuations can in fact lead to the reentrance in the flow curve responsible of a discontinuous yielding transition. 

The discontinuous yielding we have obtained occurs for vales of $B/\mu$ that are quite small, less than $\sim 0.15$. This seems to place this effect in a rather theoretical limit, without many practical implications (note in particular that the effect will occur in materials with negative values of Poisson ratio). However we want to stress that this maximum value of $B/\mu$ for the effect to occur may drastically increase if some other parameters in the problem are changed. 
In our overdamped implementation of the model one such parameter is the ratio between the effective viscosity of the shear mode and that of the volume mode, namely the ratio $\lambda_1/\lambda_0$. In all the presentation we have kept this ratio equal to 1. However,
as shear and volume modes are qualitatively different, there is no reason to stick to this value. We have observed that the maximum value of $B/\mu$ to observe discontinuous yielding and stable shear bands formation increases strongly with $\lambda_1/\lambda_0$. For instance, the case $B/\mu=1$ displays discontinuous yielding if $\lambda_1/\lambda_0=10$. 
Therefore, although the present analysis was done in the context of a model that assumes an overdamped dynamics that may not be quite realistic in its application to concrete practical cases, it is likely that larger values of $B/\mu$ can produce a first order yielding if other parameters are appropriately tuned, even in the absence of more specific effects such as aging stabilization.

\section{Acknowledgments}
I thank Ezequiel Ferrero for helpful stimulating discussions.


\begin{thebibliography}{1}
\bibitem{foot}Note that although this scenario may look similar to the flow-concentration coupling mentioned in the previous paragraph, we will not actually introduce a coupling between shear and volume fluctuations (in particular, the critical stress of the material will be assumed to be independent of the local density), but simply we will model the fact that the bulk modulus is finite. The introduction of a shear and volume coupling in the context of the present numerical modelling will be presented elsewhere.
\bibitem{nota2}Note that expression (\ref{sigma2}) is slightly different from the one corresponding to a normal PT model described by Equation (\ref{ttilde}). In that case the value of $\sigma$ would be
$\sigma_{PT}= \mu(\dot\gamma \tilde t-e_2)/2.$
\end{thebibliography}
\end{document}